\documentstyle[12pt]{article}
\begin{document}
 
\newcommand{\oeawpreprint}[2]
{
\noindent
\begin{minipage}[t]{\textwidth}
\begin{center}
\framebox[\textwidth]{$\rule[6mm]{0mm}{0mm}$ 
\raisebox{1.3mm}{Institut f\"ur Hochenergiephysik der \"Osterreichischen
Akademie der Wissenschaften}}

\vspace{2mm}    \rule{\textwidth}{0.2mm}\\
\vspace{-4mm}   \rule{\textwidth}{1pt}
\mbox{ }    #1    \hfill    #2   \mbox{ }\\
\vspace{-2mm}   \rule{\textwidth}{1pt}\\
\vspace{-4.2mm} \rule{\textwidth}{0.2mm}
\end{center}
\end{minipage}

}   
\hyphenation{mul-ti-pli-ci-ty}
\hyphenation{ana-ly-sis}
\hyphenation{mul-ti-no-mials}
\hyphenation{mul-ti-no-mial}


\oeawpreprint{December 1995}{HEPHY-PUB 633/95}

\vspace*{5mm}
\mbox{ }\hfill hep-ph/9604373 

\vspace*{20mm}

\begin{center}
\baselineskip 18pt
{\Large\bf
Generalized moments and cumulants  \\
for samples of fixed multiplicity} \\

\baselineskip 14pt
\mbox{ }\\
P.\ Lipa, H.C.\ Eggers\footnote{
             Present address: Department of Physics, University of
             Stellenbosch, 7600 Stellenbosch, South Africa.}
and B.\ Buschbeck \\
\mbox{ }\\
{\it 
     Institut f\"ur Hochenergiephysik der \"Osterreichischen
     Akademie der Wissenschaften,} \\
{\it 
     Nikolsdorfergasse 18, A--1050 Vienna, Austria}\\
\end{center}

\vspace*{1.9cm}

\begin{abstract}\noindent
Factorial moments and cumulants are usually defined with respect to the
unconditioned Poisson process.  Conditioning a sample by selecting
events of a given overall multiplicity $N$ necessarily introduces
correlations.  By means of Edgeworth expansions, we derive generalized
cumulants which define correlations with respect to an arbitrary
process rather than just the Poisson case.  The results are applied to
correlation measurements at fixed $N$, to redefining short-range
vs.\ long-range correlations and to normalization issues. \\
\mbox{ }\\  \noindent
{PACS 13.85.Hd, 02.50.-r, 05.40.+j, 25.75.Gz}
\end{abstract}
 
\vspace*{20mm}
\centerline{To appear as Phys.\ Rev.\ D {\bf 53}, 4711 (1996).}

\newpage

In high energy multiparticle production, the variation of correlations
with overall multiplicity $N$ is most sensitive to the underlying
dynamics \cite{Sin74a,Foa75a,Bel84a,UA5-88a,UA1-90b}.  This sensitivity
provides a good opportunity to test and discriminate between various
models. However, with the notable exceptions of
Refs.\ \cite{Ber75a,Car91d}, little attention has so far been devoted
to a systematic treatment of $N$-dependent correlation measures.
\\

Selecting events by fixed values of $N$ is an example of {\it
conditioning}. One of the merits of this type of conditioning is the
separation of so-called ``short-range''  from ``long-range''
correlations.  The latter arise from non-poissonian overall
multiplicity distributions as well as the $N$-dependence of the
conditional single particle density $\rho(x|N)$  (where $x$ is any
phase space variable of interest).
\\

Every conditioning, however, introduces new correlations which can be
regarded as ``unphysical'' since they are merely a consequence of the
selection procedure. Consequently, the very concept of a correlation
must be rethought and generalized for the case of conditioned samples.
\\

In this paper, we seek to clarify and generalize 
factorial moments and cumulants for the case of fixed-multiplicity
samples. This generalization suggests a natural strategy for separating
nontrivial fixed-$N$ correlations from overall multiplicity
conditioning effects.  Generalized cumulants will hence also be useful
in the study of pion interferometry as a function of $N$
and other situations requiring conditioning.
\\

Multiparticle final states are best understood in terms of {\it point
processes} \cite{Sri74a,Egg93a}.  The concept of correlation in
point processes is usually based on the fundamental {\it Poisson
process}, which is characterized by the factorial moment generating
functional (f.m.g.f\,l.)
\begin{equation}
\label{pzc}
Q_\gamma[\lambda(x)] =
\exp\left[ - \int \lambda(x) \rho(x) \, dx \right] \,,
\end{equation}
where $\rho(x) = (1/\sigma_I)d\sigma_{\rm incl}/dx$ is the differential
inclusive cross-section at point $x$.  Equivalently, for the Poisson
process every finite-dimensional {\em counting distribution\/}, i.e.\
the joint probability of the particle counts ${\vec
n}=(n_1,\ldots,n_M)$ in an arbitrary finite partition of an overall
phase space domain $\Omega_{\rm tot}$ into $M$ nonoverlapping subdomains
(``bins'') ${\vec\Omega}=(\Omega_1,\ldots,\Omega_M)$, factorizes into a
product of $M$ Poissonians
\begin{equation}
\label{pzd}
\gamma({\vec n};{\vec\nu}) = 
\prod_{m=1}^M \gamma(n_m; \nu_m) \ ,
\end{equation}
where $\gamma(n_m; \nu_m) = \exp(-\nu_m) \nu_m^{n_m} / n_m!$  
with mean multiplicity  $\nu_m = \int_{\Omega_m} \rho(x) dx$.
\\

Particles created in a Poisson process are completely uncorrelated
in the sense that all factorial cumulant densities of order $q > 1$  
vanish,
\begin{equation}
\label{pze}
\kappa_q(x_1,\ldots,x_q) \equiv \prod_{i=1}^q
\left( -\delta \over \delta\lambda(x_i) \right) \ln Q_\gamma[\lambda(x)] 
\biggr|_{\lambda = 0}
= 0 \,,
\end{equation}
while $\kappa_1(x) = \rho(x)$. 
\\

By contrast, the {\it multinomial process\/} is defined by placing
a fixed number $N$ of particles {\em independently\/} of each other into
the domain $\Omega_{\rm tot}$ according to a probability density
$p(x|N)=\rho(x|N)/N$; it has a f.m.g.f\,l.
\begin{equation}
\label{pzf}
Q_N^{\mbox{\scriptsize mult}}[\lambda(x)] = 
\left[ 1 - \int dx \, \lambda(x) \frac{\rho(x|N)}{N}  \right]^N \,,
\end{equation}
and its finite-dimensional counting distributions are multinomials.
\\

Despite the fact that the particles are placed independently of each
other, the standard factorial cumulants of the multinomial process are
nevertheless nonzero, thereby indicating purely ``external''
correlations:
\begin{equation}
\label{pzg}
\kappa_q^{\mbox{\scriptsize mult}} (x_1,\ldots,x_q | N) = 
   (q-1)! \; (-N)^{1-q} 
   \prod_{i=1}^q \rho(x_i|N) \ .
\end{equation}
In other words, standard cumulants measure not only the ``internal''
correlations between particles but also the
deviation of the overall multiplicity $N$ from the Poisson distribution
$\gamma(N;\bar N) = \exp(-\bar N) {\bar N}^N / N!$.
\\

It thus seems natural to ask for quantities that are sensitive
not to such ``external'' deviations from the overall Poissonian
but only to ``internal'' correlations, i.e.\ deviations from
the multinomial.
\\

The answer, we believe, lies in the use of {\it generalized cumulants}
which successfully separate the two. To justify their existence and
form, we turn to the alternative definition of cumulants as coefficient
functions in an Edgeworth expansion.  In most of the literature, the
latter is understood as the expansion of a continuous univariate
distribution $P(x)$ around the gaussian (cf.\ \cite{Stu87a}).  For
integer counts, it is more natural to define a multivariate {\it
discrete Edgeworth expansion} around the Poisson process. In its most
general form, the discrete Edgeworth expansion of a point process $f$
with respect to the Poisson process $\gamma$ reads, in terms of their
f.m.g.f\,l.'s
\begin{eqnarray}
\label{pzh}
Q_f[\vec\lambda] &=& 
\exp\left[
   - \int dx [ \kappa_1^f(x) - \rho(x)] \lambda(x) 
    \right. \nonumber \\
&& \left. + \sum_{q=2}^\infty {(-1)^q \over q!}    
     \int d\vec x \, \kappa_q^f(\vec x) \vec\lambda(\vec x)
 \right]
Q_\gamma[\vec\lambda]  \,, \nonumber \\
&=& 
\exp\left[ \sum_{q=1}^\infty {(-1)^q \over q!}    
     \int d\vec x \, \kappa_q^f(\vec x) \vec\lambda(\vec x)
 \right]
\end{eqnarray}
with $\vec x = (x_1,\ldots,x_q)$ and 
$\vec\lambda(\vec x) = \lambda(x_1)\cdots\lambda(x_q)$.
The univariate expansion of the overall multiplicity distribution
$f(N)$ in terms of the Poissonian is
\begin{equation}
\label{pzi}
f(N) = \exp\left[ (\kappa_1^f - \bar N) (-\nabla)
+ \sum_{q=2}^\infty {\kappa_q^f \over q!} (-\nabla)^q
\right]   \gamma(N;\bar N) \,,
\end{equation}
where $\nabla$ is the discrete difference operator such that
$\nabla\gamma(n;\nu) \equiv \gamma(n;\nu) - \gamma(n-1;\nu)$.
\\

The extension to a  multivariate expansion for the joint counting
distributions in arbitrary finite partitions of $\Omega_{\rm tot}$ 
then reads
\begin{equation}
\label{pzj}
f({\vec n}) =
\exp\left[
  \sum_{m=1}^M [\kappa_1^f(m) - \nu_m ](-\nabla_m) 
+ \sum_{q=2}^\infty {(-1)^q \over q!}
    \sum_{m_1,\ldots,m_q} \kappa_q^f({\vec m}) 
    \nabla_{m_1}\cdots \nabla_{m_q}
\right]
\gamma({\vec n};{\vec\nu}) 
\end{equation}
where
$\nabla_{m} \gamma({\vec n};{\vec\nu})
\equiv 
\gamma({\vec n};{\vec\nu}) 
  - \prod_{m^\prime} 
    \gamma(n_{m^\prime} {-} \delta_{m^\prime m}; {\vec\nu})$,
and 
$\kappa_q^f({\vec m})$ is the $q$-fold integral
of $\kappa_q^f(\vec x)$ over the domains $\Omega_{m_1},\ldots,
\Omega_{m_q}$.
\\

It can be verified easily that the coefficients $\kappa_q^f$ in these
expansions are identical with the $q$th order multivariate factorial
cumulants of the process $f$.  Therefore standard cumulants can be {\em
defined\/} as the coefficient functions of the Edgeworth expansion
around the Poisson process.
\\

This definition can be generalized in an obvious way.
From Eq.\ (\ref{pzh}), it is  easy to show that
the generating functional $Q_f$ of a process $f$ can be
expanded in terms of the  f.m.g.f\,l.\ of {\em any other\/} 
process $h$,\footnote{
It is understood that such relations are valid only for processes
having well-defined cumulants of all orders.}
\begin{equation}
\label{pzk}
Q_f[\vec\lambda] = \exp\left[
\sum_{q=1}^\infty {(-1)^q \over q!}    \int d\vec x \, 
[\kappa_q^f(\vec x) - \kappa_q^h (\vec x)]
\vec\lambda(\vec x) \right]
Q_h[\vec\lambda]  \,,
\end{equation}
where again $\kappa_q^f$ and $\kappa_q^h$ represent the
standard factorial cumulants of $f$ and $h$. 
\\

It is quite natural, then, to define {\it generalized cumulants}
as the coefficients in the expansion of $f$ in terms of $h$,
\begin{equation}
\label{pzl}
\kappa_q^{f{/}h} \equiv \kappa_q^f - \kappa_q^h
\end{equation}
for $q \ge 2$, while for $q{=}1$, we set by convention
$\kappa_1^{f{/}h} \equiv \kappa_1^f$.  If $h$ is the Poisson process,
these generalized cumulants reduce to the standard cumulants.
\\

The generalized cumulants can be found as  functional derivatives of
the {\it ratio\/} of generating functionals,
\begin{equation}
\label{pzm}
\kappa_q^{f{/}h} = \prod_{i=1}^q 
\left( -\delta \over \delta\lambda(x_i) \right)
\ln \left( Q_f Q_\gamma \over Q_h \right)
\biggr|_{\lambda = 0} \,.
\end{equation}
where the Poisson generating functional $Q_\gamma$ has to satisfy
$(\delta Q_\gamma / \delta\lambda)|_{\lambda = 0}  = \kappa_1^h$. 
This ensures that the above convention for
$\kappa_1^{f/h}$ is satisfied\footnote{
One can view the ratio $Q_h/Q_\gamma$ as the {\it central moment\/}
generating functional of the process $h$.
}.
Note that $\kappa_q^{f{/}h}$ is  generally a standard cumulant to some
well defined process only if the process $f$ is divisible by $h$ in the
sense that the ratio $Q_f/Q_h$  itself again satisfies all requirements
for a f.m.g.f\,l.
\\

The general validity of Eq.\ (\ref{pzk}) means that
generalized cumulants will work for any type of conditioning,
i.e.\ show up the real correlations of a process $f$
with respect to a given conditioned reference process $h$.
\\

We now apply the concept of generalized cumulants to fixed-$N$
multiplicity samples. The discussion will range from the definition of
internal cumulants for fixed-$N$ measurements to resulting cumulant
formulae for fixed-bin and correlation integral measurements, followed
by a short discussion of improved short-range vs.\ long-range
correlation formulae,  summing over windows in $N$, internal moments,
and normalization.
\\

{\it Internal cumulants\/} are the generalized cumulants with 
respect to the multinomial process (\ref{pzf}):
\begin{equation}
\label{qsu}
\kappa_q^I(\vec x|N) \equiv \kappa_q(\vec x|N) 
- \kappa_q^{\mbox{\scriptsize mult}}(\vec x|N) \,,
\end{equation}
with $\kappa_q^{\mbox{\scriptsize mult}}(\vec x|N)$ given by
Eq.\ (\ref{pzg}) and $\kappa_q(\vec x|N)$ the standard cumulant
measured for a fixed-$N$ process.  These internal cumulants will vanish
everywhere, even on a differential level, if the $N$ particles in the
full domain $\Omega_{\rm tot}$ are placed independently of each other
following the same probability density $p(x|N)$, i.e.\ behave
multinomially.
\\

Univariate internal cumulants are obtained from the 
multivariate ones by integrating all $q$ variables $x_1,\ldots,x_q$
over the same domain $\Omega_m$, $\kappa_q^I(m | N) \equiv
\int_{\Omega_m} d\vec x \,\kappa_q^I(\vec x|N)$, yielding
\begin{eqnarray}
\label{qsv}
\kappa_2^I(m|N) &=& \langle n_m^{[2]} \rangle_N -
                  ( 1 - N^{-1}) \langle n_m \rangle_N^2 \,,
\nonumber\\
\kappa_3^I(m|N) &=& \langle n_m^{[3]} \rangle_N 
- 3 \langle n_m^{[2]} \rangle_N \langle n_m \rangle_N
+ 2 (1 - N^{-2}) \langle n_m \rangle_N^3 \,,
\nonumber
\end{eqnarray}
etc., with $n^{[q]} = n!/(n{-}q)!$. Event averages $\langle\rangle_N$
are taken over subsamples of fixed $N$.  These integrated internal
cumulants differ from the inclusive ones \cite{Car90d} by a correction
factor $(1-N^{1-q})$ in the last term.
\\

Integrated internal cumulants of order $q\ge 2$ vanish {\it for any
fixed-$N$ distribution} if the integration domain $\Omega_m$ is
enlarged to the full domain $\Omega_{\rm tot}$,
in contrast to the standard cumulants which integrate to
$(-1)^{q-1} \, (q-1)! \, N$.
\\

The same procedure can be used in the correlation integral method. In
the notation of Ref.\ \cite{Egg93a}, if $a \equiv \sum_j \Theta(\epsilon -
X_{ij})$ is the ``sphere count'' around particle $i$ in event $a$ and
$\langle b \rangle_N \equiv \langle \sum_j \Theta(\epsilon - X_{ij}^{ab})
\rangle_N$ is the average of counts within other events $b$, the
lowest-order internal cumulants are correspondingly given by
\begin{eqnarray}
\kappa_2^I(\epsilon|N) 
&=& \left\langle
\sum_i [ a - (1 - N^{-1}) \langle b \rangle_N ] 
    \right\rangle_{\!\!\!N} \,, \\
\kappa_3^I(\epsilon|N) 
&=& \left\langle
\sum_i [ a^{[2]} - \langle b^{[2]} \rangle_N - 2 a \langle b \rangle_N
+ 2 (1 - N^{-2}) \langle b \rangle_N^2 ] 
    \right\rangle_{\!\!\!N} \!\!\! ,
\end{eqnarray}
and so on; the same $(1 - N^{1-q})$ factors appear as coefficients of 
the last terms $\langle b \rangle_N^{q-1}$ for higher-order
internal cumulants. Again, these internal correlation integrals
vanish for distances $\epsilon$ large enough
to encompass the full domain $\Omega_{\rm tot}$.
\\

Clearly, the multinomial correction will be most important for
small $N$ and lowest order. The correction may or may not be 
substantial, depending on the numerical size of the uncorrected
$\kappa_q$. 
\\

When experimental samples are not very large, fixed-$N$ correlation
measurements become problematic due to large sampling errors. 
Therefore one traditionally sums up all fixed-$N$
cumulants \cite{Bia73a,Foa75a,UA5-88a}
to yield what are (inaccurately) called 
``short-range correlations'' (SRC); in second order
\begin{equation}
\label{qsw}
\kappa_2^{\rm SRC}(x_1,x_2) \equiv \{\kappa_2(x_1,x_2|N)\}
= \sum_N P_N \, \kappa_2(x_1,x_2|N) \,.
\end{equation}
Inclusive cumulants are then split up into the SRC contribution plus
``long-range correlations'' (LRC):
\begin{equation}
\label{qsy}
\kappa_2(x_1,x_2) = \{ \kappa_2(x_1,x_2|N) \}
+ \{ \Delta \rho_N(x_1) \Delta\rho_N(x_2) \} \,,
\end{equation}
where $\Delta \rho_N(x) = \rho_1(x|N) -  \{\rho_1(x|N)\}$.
The SRC sum over fixed-$N$ cumulants is purported
to represent an average over the standard fixed-$N$ correlations, while
the second term represents the LRC contributions resulting from the
strong variation of $\rho(x|N)$ with overall multiplicity $N$.
\\

As we have shown, it is preferable to replace the $\kappa_q(\vec x|N)$
by internal cumulants $\kappa_q^I(\vec x|N)$ to eliminate the fixed-$N$
multinomial contributions from the short range part. We therefore
propose that the traditional SRC/LRC formula should be modified in
favor of splitting the inclusive cumulants into $N$-averages over
``internal'' and ``external'' correlations. In second order, this would
be
\begin{eqnarray}
\label{qsz}
\kappa_2(x_1,x_2) &=& \{ \kappa_2^I(x_1,x_2|N) \} \\
&+& \{ \kappa_2^{\mbox{\scriptsize mult}} (x_1,x_2|N)
     + \Delta \rho_N(x_1) \Delta\rho_N(x_2) \} \,. \nonumber
\end{eqnarray}
LRC/SRC formulae up to fourth order were catalogued in Ref.\
\cite{Car91d}. In all cases, ``internal'' correlations would be given by
$\{\kappa_q(\vec x|N) - \kappa_q^{\mbox{\scriptsize mult}}(\vec x|N) \}$,
while the ``external'' correlations correspond to the formulae in
\cite{Car91d} plus the appropriate multinomial cumulant.
\\

For practical reasons, $N$-averages of internal correlation
measures over limited multiplicity ranges $[A,B]$
are of importance,
\begin{equation}
\label{qsa}
\kappa_q^I(\vec x|A,B) = 
{ \sum_{N=A}^B P_N\, \kappa^I_q(\vec x | N)  \over 
  \sum_{N=A}^B P_N  } \,.
\end{equation}

Fixed-$N$ correlations are but one example of generalized cumulants.
Equation (\ref{pzk}) shows that subtraction of cumulants is appropriate
when {\it any} kind of reference distribution is given. Applications
which come to mind immediately are subensembles characterized by a
given number of jets and any Monte Carlo-generated simulation.
Differences between real data and the dynamics contained in a given
Monte Carlo code would be quantifiable again by a generalized cumulant
$\kappa\equiv \kappa_{\mbox{\scriptsize data}} - \kappa_{MC}$.
\\


{\it Generalized moments} $\rho_q^{f{/}h}$ are defined as
functional derivatives of the ratio of generating functionals,
\begin{equation}
\label{itm}
\rho_q^{f{/}h}(\vec x) = \prod_{i=1}^q 
\left( -\delta \over \delta\lambda(x_i) \right)
\left( Q_f Q_\gamma \over Q_h \right)
\biggr|_{\lambda = 0} \,;
\end{equation}
for the fixed-$N$ case, the corresponding internal moments are
related to the standard moments by
\begin{eqnarray}
\label{itn}
\rho_1^I(x|N) &=& \kappa_1^I(x|N) = \rho_1(x|N) \,, \\
\rho_2^I(x_1,x_2|N) 
&=& \kappa_2^I(x_1,x_2|N) + \kappa_1^I(x_1|N)\kappa_1^I(x_2|N) 
    \\
&=& \rho_2(x_1,x_2|N) + (1/N)\, \rho_1(x_1|N) \, \rho_1(x_2|N)
    \,, \nonumber
\end{eqnarray}
and further, omitting the arguments for brevity,
\begin{eqnarray}
\rho_3^I
&=& \rho_3 + \rho_1 \rho_1 \rho_1
      \left( {3\over N} - {2\over N^2} \right)
    \,, \\
\rho_4^I
&=& \rho_4 + {1\over N}\sum_{(6)}\rho_1\rho_1\rho_2 
   - \rho_1 \rho_1 \rho_1 \rho_1 \left( {5\over N^2} - {6\over N^3} \right)
\,,
\end{eqnarray}
the brackets under the sum indicating the number of permutations.
Internal moments have the property that when the
{\it measured\/} moment $\rho_q(\vec x|N)$ behaves multinomially,
$\rho_q(\vec x|N) 
\stackrel{\rm mult}{\longrightarrow} (N^{[q]}/N^q)\, 
                       \prod_{i=1}^q\rho_1(x_i|N)
$,
they factorize without a prefactor:
\begin{equation}
\label{ito}
\rho_q^I(\vec x|N) 
\stackrel{\rm mult}{\longrightarrow} \prod_{i=1}^q\rho_1(x_i|N) \,.
\end{equation}

For a subsample of fixed $N$, moments and cumulants at fixed
multiplicity can be normalized in two ways. The most natural
normalization procedure is to use in the denominator exactly that
quantity which the numerator would default to if the process were fully
independent. Hence the normalization for fixed-$N$ factorial moments
over some domain $\Omega_m$ would be
\begin{eqnarray}
\label{itp}
F_q(\Omega_m|N) &=&
{\int_{\Omega_m}  d\vec x \, \rho_q(\vec x|N) \over
\int_{\Omega_m}  d\vec x \, \rho_q^{\rm mult}(\vec x|N)  }
\nonumber \\
&=&
{N^q \over N^{[q]} }\;
{\int_{\Omega_m}  d\vec x \, \rho_q(\vec x|N) \over
\int_{\Omega_m}  d\vec x \,  \rho_1(x_1|N) \cdots \rho_1(x_q|N) } \,,
\end{eqnarray}
while the internal moment would be normalized according to
\begin{equation}
\label{itq}
F_q^I(\Omega_m|N) = 
{\int_{\Omega_m}  d\vec x \, \rho_q^I(\vec x|N) \over
\int_{\Omega_m}  d\vec x \,  \rho_1(x_1|N) \cdots \rho_1(x_q|N) } \,.
\end{equation}
Both these definitions yield $F_q = F_q^I \equiv 1$ for any $\Omega_m$ when
the measured $\rho_q(\vec x|N)$ behaves multinomially within the total
window $\Omega_{\rm tot}$. For internal cumulants, the same
normalization (\ref{ito}) as for the internal moments would be
appropriate.
\\

Internal moments integrate under the total phase space domain 
$\Omega_{\rm tot}$ to
$ \int_\Omega d\vec x \, \rho_q^I(\vec x) = N^q$,
while $\rho_q(\vec x|N)$ integrates to $N^{[q]}$, so that both
normalized moments become unity when integrated over
$\Omega_{\rm tot}$ for {\it any\/} distribution: 
\begin{equation}
\label{itr}
F_q^I(\Omega_{\rm tot} | N) = F_q(\Omega_{\rm tot} | N) = 1 \,.
\end{equation}
The integral over
$\Omega_{\rm tot}$ of normalized internal cumulants will, of course, be
zero just as the unnormalized ones.
\\

In summary, we have shown that generalized cumulants are an improved
measure of correlations in samples of fixed multiplicity.  The latter
are more sensitive in discriminating between dynamical mechanisms and
models than inclusive quantities.  The procedure outlined here can be
applied to a number of other conditioning problems within and beyond
high energy physics.  Detailed applications to correlation measurements
such as pion interferometry are in progress.
\\

{\bf Acknowledgements} \\

We wish to thank Peter Carruthers for encouraging discussions. 
PL gratefully acknowledges support by the Austrian Academy of Sciences 
through an APART (Austrian Programme for Advanced Research and
Technology) fellowship. HCE was kindly supported by a Lise-Meitner
fellowship of the Austrian Fonds zur F\"orderung der Wissenschaftlichen
Forschung (FWF).
\\

\newpage



\begin{thebibliography}{99}

\bibitem{Sin74a}
         R.\ Singer et al., Phys.\ Lett.\ {\bf 49B}, 481 (1974);
         K.\ Eggert et al., Nucl.\ Phys.\ {\bf B86}, 201 (1975);
         B.\ Y.\ Oh et al., Phys.\ Lett.\ {\bf 56B}, 400 (1975).

\bibitem{Foa75a}L.\ Fo\`a, Phys.\ Rep.\ {\bf 22}, 1 (1975).

\bibitem{Bel84a}W.\ Bell et al., Z.\ Phys.\ {\bf C22}, 109 (1984);
                {\bf C32}, 335 (1986).

\bibitem{UA5-88a}UA5 Collaboration, R.\ E.\ Ansorge et al., 
                 Z.\ Phys.\ {\bf C37}, 191 (1988).

\bibitem{UA1-90b}UA1 Collaboration, C.\ Albajar et al., 
            Nucl.\ Phys.\ {\bf B345}, 1 (1990);
           UA1 Collaboration, Y.F.\ Wu et al.,
           Act.\ Phys.\ Slov.\ {\bf 44}, 141 (1994);
           P.\ Lipa, B.\ Buschbeck and H.C. Eggers, to be published.

\bibitem{Ber75a}E.\ L.\ Berger, Nucl.\ Phys.\ {\bf B85}, 61 (1975).

\bibitem{Car91d}P.\ Carruthers, Phys.\ Rev.\ A{\bf 43}, 2632 (1991).

\bibitem{Sri74a}S.K.\ Srinivasan, 
           {\it Stochastic Point Processes}, Griffin's Statistical
           Monographs and Courses Vol.\ 34, Griffin (1974).

\bibitem{Egg93a}H.\ C.\ Eggers, P.\ Lipa, P.\ Carruthers and B.\ Buschbeck, 
           Phys.\ Rev.\ D{\bf 48}, 2040 (1993).

\bibitem{Stu87a}A.\ Stuart and J.\ K.\ Ord, 
           {\it Kendall's Advanced Theory of Statistics}, 
           Vol.1, fifth edition, Oxford University Press,  New York (1987).

\bibitem{Car90d}P.\ Carruthers, H.C.\ Eggers, and I.\ Sarcevic,
           Phys.\ Lett.\ {\bf 254B}, 258 (1991).

\bibitem{Bia73a}A.\ Bia\l as, in: {\it 4th International Symposium
           on Multiparticle Dynamics}, Pavia, Italy, 1973,
           edited by F.\ Duimio, A.\ Giovannini and S.\ Ratti
           World Scientific (1974).

\end{thebibliography}
\end{document}